\documentclass[smallextended]{svjour3}
\usepackage[T1]{fontenc}
\usepackage[latin9]{inputenc}
\usepackage{wrapfig}
\usepackage{url}
\usepackage{graphicx}

\makeatletter
\RequirePackage{fix-cm}

\smartqed  
\usepackage{sidecap}
\makeatother

\begin{document}

\title{Effects of \textsuperscript{4}He film on quartz tuning forks in
\textsuperscript{3}He at ultra-low temperatures\thanks{}}

\titlerunning{Effects of \textsuperscript{4}He film on quartz tuning forks in
\textsuperscript{3}He at ultra-low temperatures}

\author{T. S. Riekki \and J. Rysti \and J. T. M\"akinen \and A. P. Sebedash \and
V. B. Eltsov \and \\ J. T. Tuoriniemi}

\authorrunning{T. S. Riekki \and  J. Rysti  \and J. T. M\"akinen \and  A. P. Sebedash \and  V. B. Eltsov
\and  J. T. Tuoriniemi}

\institute{T. S. Riekki$^\dagger$ \and  J. Rysti \and J. T. M\"akinen \and  V. B. Eltsov \and  J. T. Tuoriniemi\at Aalto
University - School of Science \\
Department of Applied Physics\\
Low Temperature Laboratory\\
P.O. BOX 15100 FI-00076 Aalto, Finland\\
$^\dagger$\email{tapio.riekki@aalto.fi}\\
A. P. Sebedash \at P. L. Kapitza Institute for Physical Problems
RAS\\
Kosygina 2, 119334 Moscow, Russia\\
}

\date{Received: date / Accepted: date}
\maketitle
\begin{abstract}
In pure superfluid \textsuperscript{3}He-B at ultra-low temperatures, quartz tuning
fork oscillator response is expected to saturate when the dissipation
caused by the superfluid medium becomes substantially smaller than
the internal dissipation of the oscillator. However, even with small amount
of \textsuperscript{4}He covering the surfaces, we have observed saturation already at significantly higher temperatures
than anticipated, where we have other indicators to prove that the
\textsuperscript{3}He liquid is still cooling. We found that this anomalous behavior has a rather strong pressure dependence,
and it practically disappears above the crystallization pressure of
\textsuperscript{4}He. We also observed a maximum in the fork resonance frequency at temperatures where the transition in quasiparticle flow from the hydrodynamic to the ballistic regime is expected. We suggest that such anomalous features derive from the superfluid \textsuperscript{4}He film on the
oscillator surface.

\keywords{Quartz tuning fork \and Helium-3 \and Helium-3\textendash Helium-4 mixture \and 
Helium-4 film}
\end{abstract}

\section{Introduction\label{sec:Introduction}}

Quartz tuning forks (QTFs) are used for temperature, pressure, viscosity and turbulence
measurements in normal and superfluid helium \cite{Blaauwgeers,Pentti_Rysti_Salmela,Bradley_Transition2turbulence,Makinen2018}. These influence the width (full width at half maximum) and the frequency of the fork resonance.
The characteristic dimensions of a typical QTF may also
match the wavelength of first or second sound in pure helium or isotope
mixtures, at certain temperature and pressure, resulting in acoustic phenomena
\cite{Forks_acoustic_phenomena,Riekki2016,Salmela_Tuoriniemi_Rysti}
that are interesting in their own right, but can also make interpreting
the fork data more difficult.

On cooling of  \textsuperscript{3}He below
the superfluid transition temperature $T_{\mathrm{c}}$, the dissipation
caused by thermal excitations, or
quasiparticles, becomes smaller, as their number decreases, which is observed as reduction in the QTF resonance width. In the B-phase at the lowest temperatures, the quasiparticle density decreases exponentially with temperature, and eventually dissipation caused by the quasiparticles becomes smaller than the internal dissipation of the fork, giving typically a residual width 10-20~mHz, which poses the low-temperature limit for thermometry.

When \textsuperscript{4}He is added to a \textsuperscript{3}He system, the fork analysis becomes more complex, as the surfaces become coated with \textsuperscript{4}He \cite{Peshkov1975,Murakawa2012}.  Below $100\,\mathrm{mK}$, the \textsuperscript{4}He layer becomes
superfluid \cite{Agnolet1989}, and due to superfluid film flow it
will spread out to cover all the surfaces of the experimental cell. The film will change the quasiparticle
reflection conditions \cite{Kim1993} on the QTF surface affecting its resonance
response. At sufficiently high pressures, the \textsuperscript{4}He
layer becomes solid, and is no longer mobile as a liquid layer would
be. However, even the presence of solid layer may affect the quasiparticle
reflection conditions, provided that the layer is thick enough to affect the surface roughness.

Boldarev \emph{et al.} \cite{Boldarev2011} observed in the \textsuperscript{3}He-rich phase of phase-separated
\textsuperscript{3}He\textendash \textsuperscript{4}He mixture, between $15$ and $350\,\mathrm{mK}$, that
the QTF deviated from the predicted viscosity and density dependent response.
They attributed this anomalous behavior to the \textsuperscript{4}He
film covering the surface of the QTF, which they estimated to have
thickness of several microns. The non-trivial response made the fork calibration more difficult, but in their experiment the fork still had clear temperature sensitivity.

We have studied the behavior of \textsuperscript{4}He-coated quartz
tuning forks in \textsuperscript{3}He at temperatures below $1\,\mathrm{mK}$, where we observed saturation in the QTFs' response at higher temperatures than anticipated. We have two independent
experiments that demonstrate similar saturation behavior:
one is a nafen-filled \textsuperscript{3}He cell with surfaces
coated with approximately 3 atomic layers of \textsuperscript{4}He,
and the other an adiabatic melting cell that contains saturated \textsuperscript{3}He\textendash \textsuperscript{4}He
mixture at \textsuperscript{4}He crystallization pressure, where we expect a much thicker equilibrium film. In both experiments we have also observed a maximum in the resonance frequency at temperatures where the quasiparticle flow regime changes from the hydrodynamic to the ballistic one. In this paper we focus on reporting our experimental observations, while the detailed explanation on the origin of the effects remains a task for the future.

\section{Results\label{sec:Results}}

\subsection{\protect\textsuperscript{3}He in phase-separated \protect\textsuperscript{3}He\textendash \protect\textsuperscript{4}He
mixture at \protect\textsuperscript{4}He crystallization pressure\label{subsec:Saturated-mixture}}
\begin{figure}
	\centering
	\includegraphics[width=0.95\linewidth]{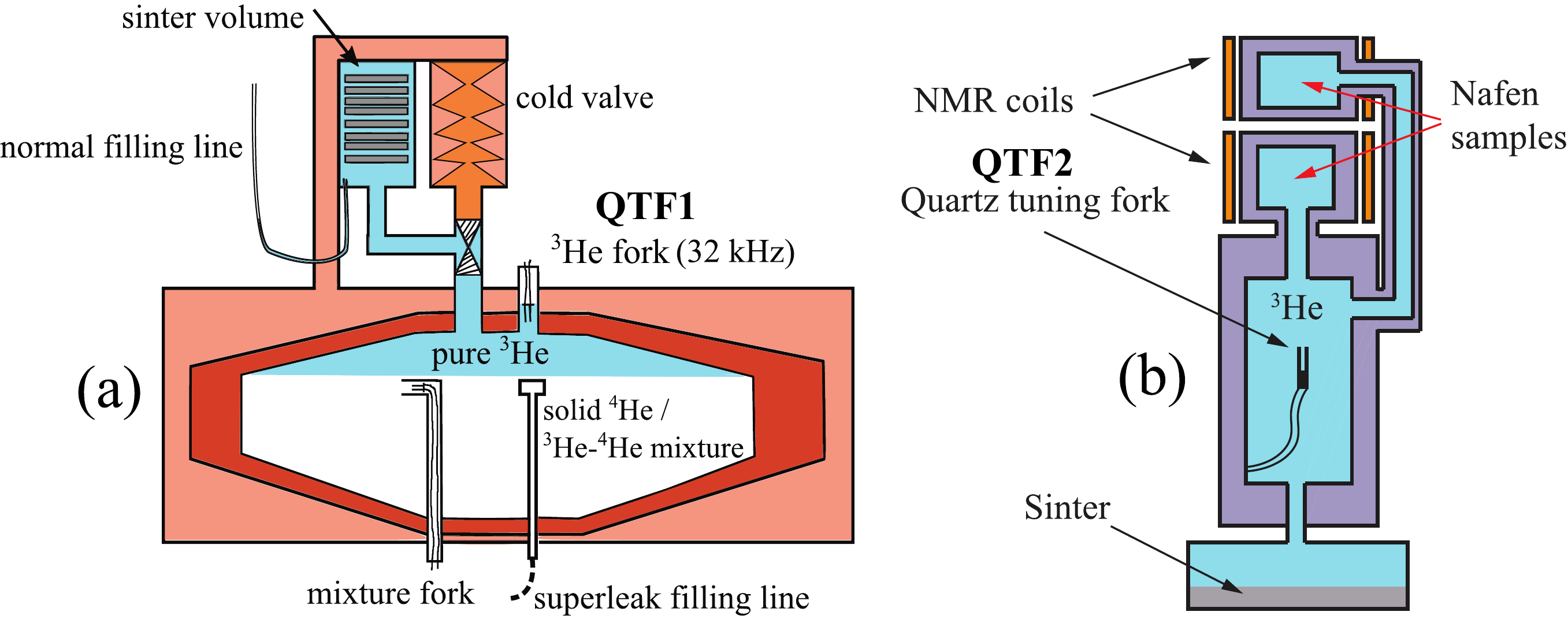}
	\caption{(color online) Schematic drawings of the experimental cells. The adiabatic melting cell (a) consists of
		a large main volume and a sinter-filled heat exchanger
		volume separated by a cold valve.
		The cell is monitored with two quartz tuning fork oscillators, one
		in the  \protect\textsuperscript{3}He phase (QTF1) and another in
		the mixture phase (or frozen in solid \protect\textsuperscript{4}He,
		depending on the stage of the experimental run). The nafen cell (b) has two separate samples with different nafen densities. They are both connected to a volume of bulk $^3$He, where the thermometer quartz tuning fork (QTF2) is located. The cell is mounted on the nuclear stage of a rotating cryostat. The surfaces in both systems are coated with a layer of $^4$He, although of different thickness.}
	\label{fig:cell-schematic}
\end{figure}
In the adiabatic melting experiment \cite{Sebedash1997,Adiabatic_Melting,Tuorinemi_Martikainen_Pentti}, sub-$0.1\,\mathrm{mK}$ temperatures in \textsuperscript{3}He--\textsuperscript{4}He mixtures are pursued
at \textsuperscript{4}He crystallization pressure $25.64\,\mathrm{bar}$ \cite{Pentti2006,Salmela2011}
by first precooling a system of  solid \textsuperscript{4}He
and  liquid \textsuperscript{3}He with an adiabatic nuclear refrigerator,
and then allowing the solid to melt by extracting \textsuperscript{4}He, mixing the two isotopes providing cooling \cite{Riekki_Thermodynamics_adiabatic}.
\begin{figure}
	\centering
	\includegraphics[width=0.95\linewidth]{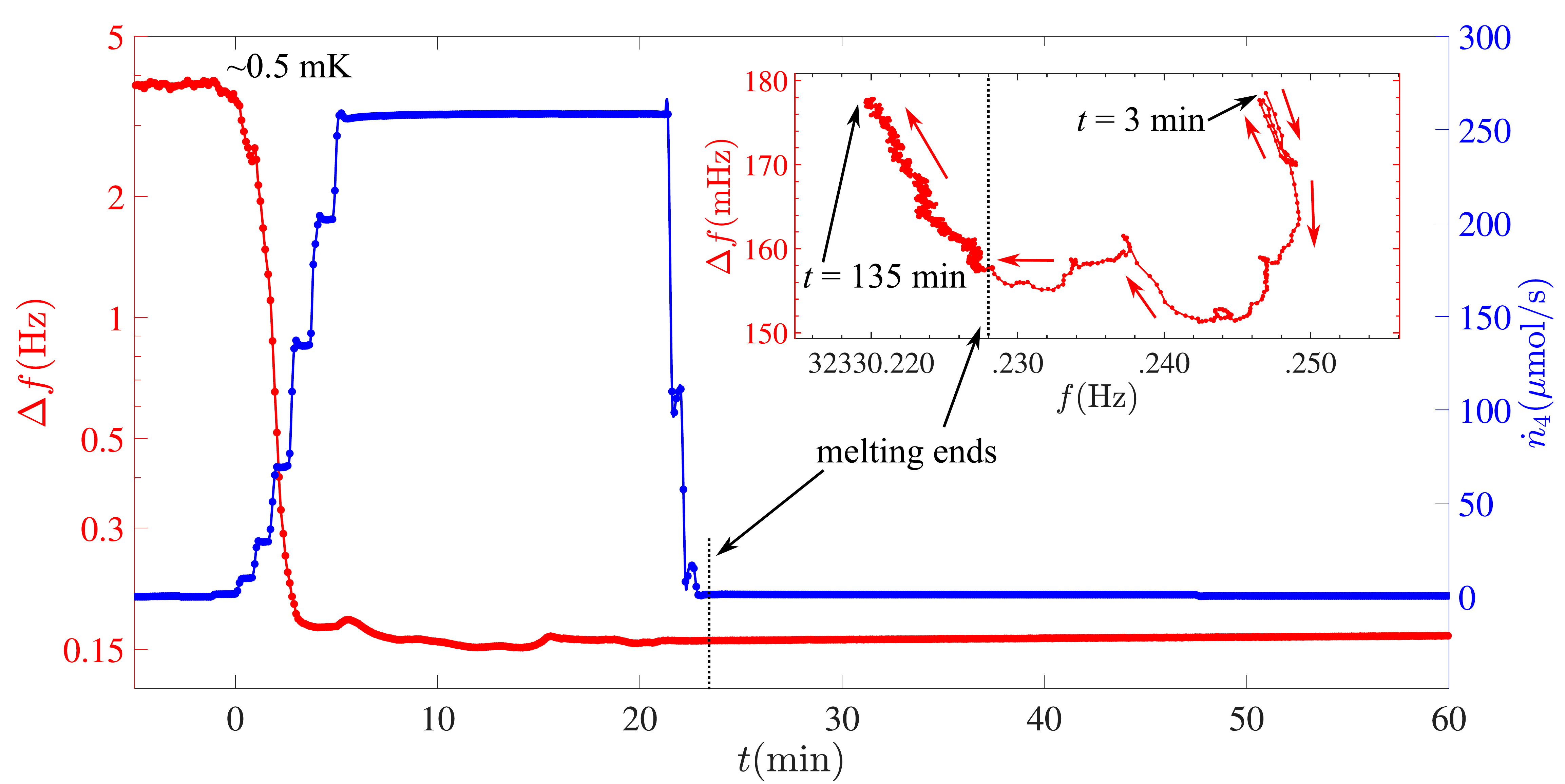}\caption{(color online) (Main panel) Left y-axis: resonance width of the quartz tuning fork (QTF1)
		in the  \protect\textsuperscript{3}He phase during the melting process with zero time chosen to be at the beginning of the
		melting. Right y-axis: \textsuperscript{4}He extraction rate. (Inset) QTF1 resonance width
		versus resonance frequency during the coldest stage of the run, showing the saturation of the width value, and the anomalous features that occur during the melting. Red arrows indicate the direction of time; at around 3 min the width backtracks slightly and then continues to decrease.\label{fig:time-width+inset} }
\end{figure}

Fig. \ref{fig:cell-schematic}a shows a sketch of the melting experiment cell. More details of this setup can
be found in Ref. \cite{Sebedash_QFS}. The resonance width of the
QTF in mixture is about $400\,\mathrm{Hz}$, and the effects of the \textsuperscript{4}He coating on its behavior are indistinguishable. On the other hand, the resonance width of the QTF in \textsuperscript{3}He (QTF1), located at the top of the main cell volume, reaches approximately $0.2\,\mathrm{Hz}$
at the end of the melting process, and then the superfluid \textsuperscript{4}He film causes it to have an anomalous response.

Fig. \ref{fig:time-width+inset} shows the QTF1 response during the melting process. The fork was measured in the tracking mode which enables us to receive datapoints every few seconds even at very narrow widths by assuming a Lorentzian lineshape with a constant area \cite{Pentti_Rysti_Salmela}.
The melting was started at around 4 Hz resonance width, corresponding to about $0.19T_\mathrm{c}\approx 0.5\,\mathrm{mK}$ temperature. Initially
the resonance width decreases rapidly as the cell cools
down, and the narrowest widths are already reached within the
first few minutes of the process. The temperature
calibration for the QTF1 was obtained using the self-calibration method
described in Ref. \cite{Todoshchenko2014}. The observed $150\,\mathrm{mHz}$ resonance width would then correspond to about
$0.11T_\mathrm{c}\approx0.3\,\mathrm{mK}$. At this temperature with the \textsuperscript{4}He extraction rate $\dot{n}_{4}\approx260~\mathrm{\mu mol/s}$, the cooling power of the melting process \cite{Riekki_Thermodynamics_adiabatic} is approximately $2\,\mathrm{nW}$. This is a much larger value than the heat leak to the cell $0.1\,\mathrm{nW}$, which was estimated during the warm-up period, after the melting, when the QTF1 width started to have temperature sensitivity again. Also, the viscous heating effects during the melting are considered insignificant. With the estimated heat leak, the liquid should cool down to below $0.3\,\mathrm{mK}$, suggesting that the resonance width is no longer proportional to the quasiparticle density in bulk.
Even after the melting was stopped, the QTF1 did not show any
rapid change from the saturation value which indicates that the
actual temperature of the liquid was lower than the value given
by the resonance width. The fork's self-calibration relies on the transition to the ballistic flow regime, the point of which cannot be determined precisely. However, we do not believe that the uncertainty in the temperature calibration could explain
the discrepancy between the temperature given by the QTF1 and the
cooling power of the melting process. Even
at $0.1\,\mathrm{mK}$, with $260\,\mathrm{\mu mol/s}$ \textsuperscript{4}He extraction rate,
the cooling power $0.2\,\mathrm{nW}$ is still larger than the estimated
external heat leak.

The inset in Fig. \ref{fig:time-width+inset} shows that during the melting, there appears anomalous features on the QTF1's frequency-width plot. These resonance-like features only occur while the melting is being carried out; they do not reproduce when the cell is slowly warming up after the melting is over. We point out that during the melting period the distance between the fork and the \textsuperscript{3}He--mixture phase-separation boundary is changing. As the melting is being carried out, \textsuperscript{3}He is dissolved into \textsuperscript{4}He released from the solid, decreasing the volume of the \textsuperscript{3}He phase, while increasing the volume of the mixture phase. When the mixture phase, containing \textsuperscript{4}He, approaches QTF1, it will increase the thickness of the \textsuperscript{4}He film covering the fork due to the Rollin film effect \cite{Rollin1,Rollin2}. Another observation, that seems to corroborate the phase-separation boundary vicinity effect, is the $30\,\mathrm{mHz}$ resonance frequency shift from the before-melting value that remained even after the melting had been stopped.

\subsection{\protect\textsuperscript{3}He with small amount of $^4$He present\label{subsec:Aerogel}}

The nafen experiment consists of two separate samples of $^3$He confined in the nematic nano-material nafen \cite{NematicAerogels}, which are connected to a volume of bulk $^3$He (Fig.~\ref{fig:cell-schematic}b). The temperature of helium is controlled by changing the magnetic field applied to the nuclear demagnetization cooling stage. The properties of $^3$He in the two nafen samples are probed by means of nuclear magnetic resonance (NMR). A quartz tuning fork in the bulk $^3$He volume (QTF2) is used as a thermometer. In this experiment $^4$He is present only to coat the surfaces of nafen to prevent the formation of paramagnetic solid $^3$He \cite{PhysRevLett.120.075301}. The thickness of $^4$He on the nafen strands was determined to be approximately 2.5 atomic layers \cite{PhysRevLett.120.075301}. The surfaces, including the quartz tuning fork, could adsorb more $^4$He, thus the $^4$He layer was not maximal \cite{PhysRevLett.120.075301}. This was clearly demonstrated after the measurements presented in this paper, as adding more $^4$He into the system and repressurizing back to 29.5~bar changed the tuning fork width at the bulk $^3$He superfluid transition from 800~Hz to 570~Hz.

\begin{figure}
	\centering
	\includegraphics[width=0.98\linewidth]{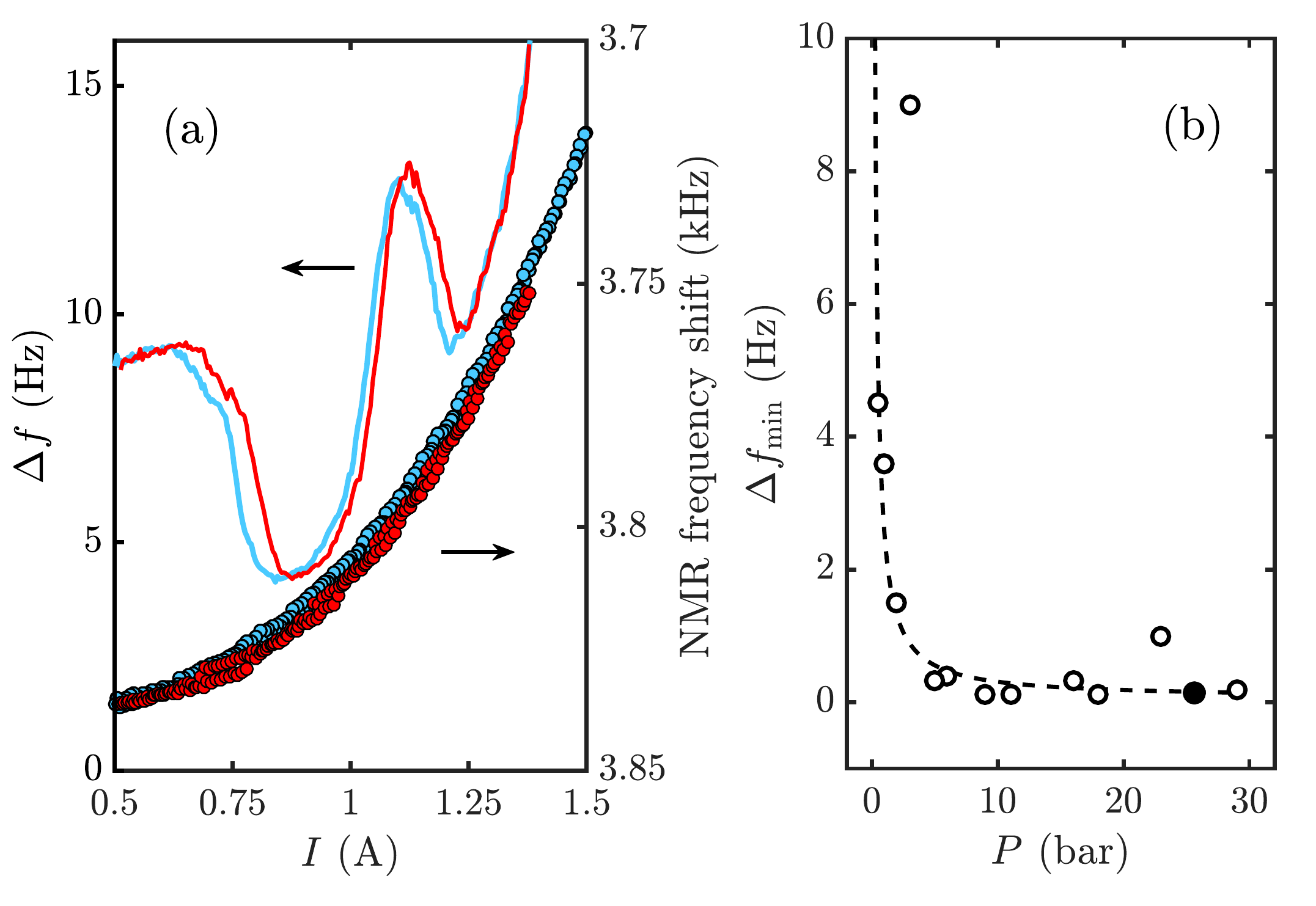}
	\caption{(color online) (a): Quartz tuning fork width (solid lines) and NMR frequency shift (dots) from the Larmor frequency (363 kHz) at 3~bar pressure as functions of the demagnetization magnet current, which controls the temperature of the sample. Blue color indicates cooling and red warming. (b): Narrowest QTF widths as a function of pressure. The dashed curve is a guide to the eye. Open points are from the nafen experiment and the black point from the adiabatic melting experiment.}
	\label{fig:width-NMR}
\end{figure}

Fig.~\ref{fig:width-NMR}a shows the QTF2 resonance width and the NMR frequency shift during cooling and warming of the sample at 3~bar pressure.
These two quantities give independent measurements of temperature. The tuning fork width displays a resonance-like feature at 1.1~A current in the demagnetization magnet, a minimum of about 4~Hz at 0.9~A, and an eventual saturation to 9~Hz toward the lowest temperatures. The NMR frequency shift, on the other hand, indicates continuous cooling of the sample all the way down to the lowest demagnetization current. The NMR frequency of superfluid $^3$He, in the polar phase confined in nafen, is shifted from the Larmor value as a function of temperature in axial magnetic field \cite{PhysRevLett.117.255301}. The QTF2 was measured by continuously sweeping over the resonance.

\begin{figure}
	\centering
	\includegraphics[width=0.98\linewidth]{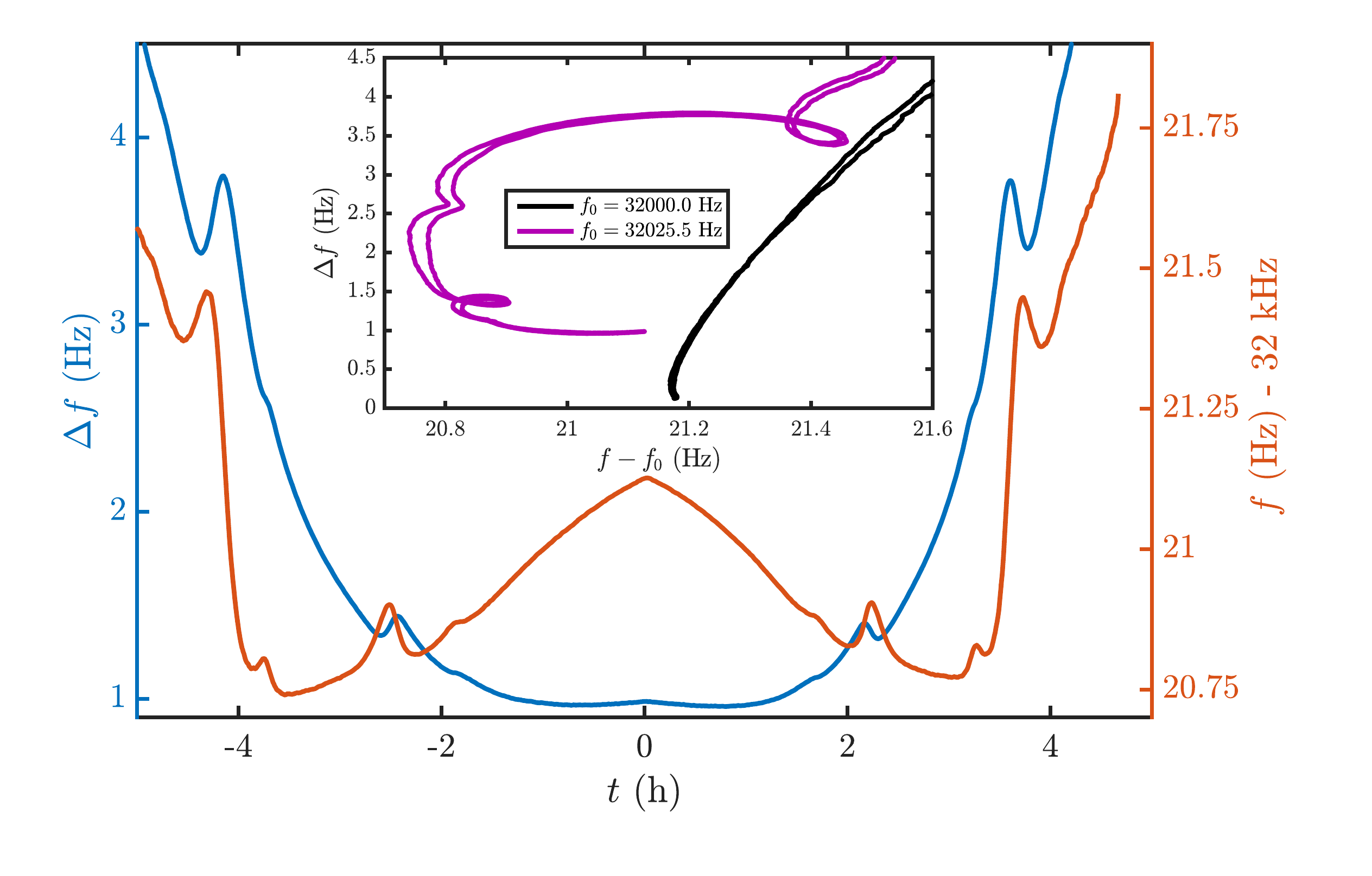}
	\caption{(color online) Quartz tuning fork width and frequency at 23~bar during cooling ($t<0$) and warming ($t>0$) of the nafen cell over a 10-hour period, demonstrating multiple resonance modes and saturation of the tuning fork width. A minimum of the width is reached at approximately $\pm 0.7$ h. The inset plots the QTF width in terms of frequency at 23~bar (red) and at 29~bar (black).}
	\label{fig:freq-width-ROTA}
\end{figure}

Fig.~\ref{fig:freq-width-ROTA} plots the QTF2 resonance width and frequency at 23~bar during slow cooling and warming.
Here the QTF2 was measured by applying pulse excitation and recording the ring-down signal. This gives superior data acquisition rate and noise at small resonance widths compared to the continuous sweeping method. Multiple resonance-like features are seen together with a shallow minimum in the width and an eventual saturation to about 1~Hz. In the absence of anomalous behavior, QTF2 width of 1~Hz would correspond to about $0.16 T_c$ temperature, or 0.4~mK at 23~bars. The frequency of the oscillator continues to change even after the width has saturated. The same pattern is repeated during warming of the sample, but the QTF2 response does not return exactly along the same path.

The anomalous behavior of the tuning fork strongly depends on pressure. It is present at 23~bars, where resonance-like features are observed and the tuning fork width would not go below 1~Hz. At 29~bar pressure, which is above the crystallization pressure of $^4$He, there is no indication of any anomalies and the tuning fork width could be reduced to 190~mHz without evidence of saturation (inset of Fig.~\ref{fig:freq-width-ROTA}). Measurements have been performed at various pressures, but the QTF2 was measured using the pulse method only at 23 and 29~bars. The continuous sweeping method may not reveal small anomaly patterns of the oscillator or the saturation. At 3~bar pressure the anomaly is the strongest. The minimum attained resonance widths as a function of pressure are plotted in Fig.~\ref{fig:width-NMR}b. Small resonance-like features are visible at 16, 6, 5, and 2~bar pressures, even with the sweeping QTF2 measurement.

The resonance-like features could be attributed to the first sound resonances in the tuning fork cavity \cite{Rysti2014}. The diameter of our tuning fork volume (9~mm) and the oscillator frequency (32~kHz) match roughly the frequencies of radial acoustic modes, especially at lower pressures, where the speed of first sound is smaller. It is not clear without more detailed analysis if the resonance-like features seen at 23~bar can be explained by acoustic resonances. The absence of these anomalies at 29 bar pressure might be due to the larger speed of sound.

\subsection{\protect QTF resonance frequency maximum \label{subsec:Ballistic}}

\begin{figure}
	\centering
	\includegraphics[width=0.95\linewidth]{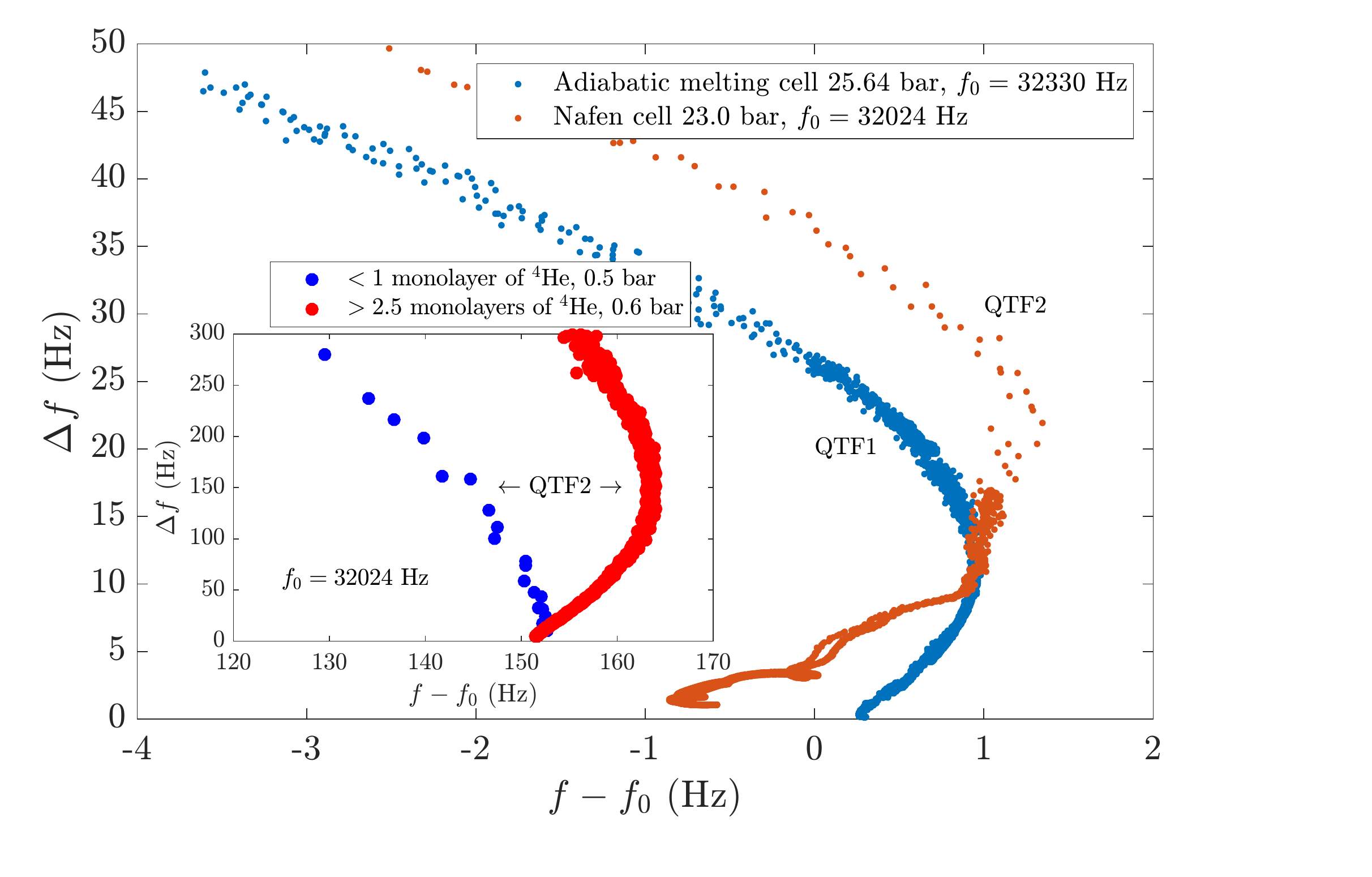}
	\caption{(color online) Main panel: Fork resonance width as a function of the resonance frequency in the adiabatic melting experiment (blue, QTF1) and in the nafen experiment (orange, QTF2), both of which show a maximum in the resonance frequency. The resonance frequencies have been shifted by $f_0$, given in the legend. The inset shows that the maximum disappears if the \textsuperscript{4}He content of the cell is low enough.}
	\label{fig:maximum}
\end{figure}

In both the used setups, we have also observed a resonance frequency maximum in the forks' response, at around 0.25$T_\mathrm{c}$. This is illustrated in the main panel of Fig.~\ref{fig:maximum}. The inset additionally shows the case of reduced \textsuperscript{4}He amount in the nafen cell. In this case, the maximum disappears, and the resonance frequency instead saturates at the lowest temperatures, which is consistent with observations in pure \textsuperscript{3}He \cite{Todoshchenko2014}.

The appearance of the frequency maximum with the increasing \textsuperscript{4}He coverage probably originates from the change in \textsuperscript{3}He quasiparticle scattering conditions on the QTF surface \cite{Kim1993}, as the thickness of the film grows. The maximum occurs at around the temperatures, in which the quasiparticle flow is expected to change from the hydrodynamic to the ballistic flow regime, and it could possibly be used as an indicator of such. Thus the maximum could be utilized in the QTF self-calibration described in Ref.~\cite{Todoshchenko2014}.

\section{Conclusions\label{sec:Conclusions}}

We have observed a saturation in the temperature dependence of the resonance width of quartz tuning fork oscillators in two independent experiments in \textsuperscript{3}He systems with surfaces coated by \textsuperscript{4}He. 
In the melting experiment, the temperature indicated by the QTF resonance width saturates to a value at which the cooling power of the helium isotope mixing process would still be significantly larger than the external heat leak. In the nafen experiment, on the other hand, we had an NMR-based thermometry that showed further cooling in the experimental cell, even after the QTF width had saturated. We also observed strong pressure dependence in the value of the saturated width, with the maximum being at $3\,\mathrm{bar}$. Both setups also displayed a maximum in the QTF resonance frequency, which appears only if there is enough \textsuperscript{4}He in the system.

We suggest that this behavior originates from the \textsuperscript{4}He film covering the QTF, however, the detailed understanding of the phenomenon requires more work. In particular, studying the dependence on the fork size might be important. As a practical conclusion, forks covered by \textsuperscript{4}He remain relatively reliable thermometers only down to widths of about 1~Hz, provided that the geometry of the fork volume excludes coupling to the acoustic modes. At pressures near the \textsuperscript{4}He crystallization pressure, this limit corresponds to approximately $0.16T_\mathrm{c}$.

\begin{acknowledgements}
This work was supported in part by the Jenny and Antti Wihuri Foundation (Grant
No. 00170320), and it utilized the facilities provided by Aalto University
at OtaNano - Low Temperature Laboratory infrastructure. We also acknowledge the support by the
European Research Council (ERC) under the European Union's Horizon 2020
research and innovation programme (Grant Agreement No. 694248).
\end{acknowledgements}

\section*{\textemdash \textemdash \textemdash \textemdash \textemdash \textemdash \textemdash{}}

\bibliographystyle{spphys}

\begin{thebibliography}{10}
	\providecommand{\url}[1]{{#1}}
	\providecommand{\urlprefix}{URL }
	\expandafter\ifx\csname urlstyle\endcsname\relax
	\providecommand{\doi}[1]{DOI \discretionary{}{}{}#1}\else
	\providecommand{\doi}{DOI \discretionary{}{}{}\begingroup
		\urlstyle{rm}\Url}\fi
	
	\bibitem{Blaauwgeers}
	R.~Blaauwgeers, M.~Blazkova, M.~{\v{C}}love{\v{c}}ko, V.~B.~Eltsov, R.~de~Graaf,
	J.~Hosio, M.~Krusius, D.~Schmoranzer, W.~Schoepe, L.~Skrbek, P.~Skyba, R.~E.~Solntsev, D.~E.~Zmeev, J. Low Temp. Phys.
	\textbf{146}, 537 (2007).
	\newblock \doi{10.1007/s10909-006-9279-4}
	
	\bibitem{Pentti_Rysti_Salmela}
	E.~Pentti, J.~Rysti, A.~Salmela, A.~Sebedash, J.~Tuoriniemi, J. Low Temp. Phys. \textbf{165}, 132 (2011).
	\newblock \doi{10.1007/s10909-011-0394-5}
	
	\bibitem{Bradley_Transition2turbulence}
	D.~I.~Bradley, M.~J.~Fear, S.~N.~Fisher, A.~M.~Gu{\'{e}}nault, R.~P.~Haley, C.~R.~Lawson, P.~V.~E.~McClintock, G.~R.~Pickett, R.~Schanen, V.~Tsepelin, L.~A.~Wheatland, J. Low Temp. Phys. \textbf{156}, 116 (2009).
	\newblock \doi{10.1007/s10909-009-9901-3}
	
	\bibitem{Makinen2018}
	J.~T.~M\"akinen, V.~B.~Eltsov, Phys. Rev. B
	\textbf{97}, 014527 (2018).
	\newblock \doi{10.1103/PhysRevB.97.014527}
	
	\bibitem{Forks_acoustic_phenomena}
	J.~Rysti, J.~Tuoriniemi, J. Low Temp. Phys.
	\textbf{177}, 133 (2014).
	\newblock \doi{10.1007/s10909-014-1203-8}
	
	\bibitem{Riekki2016}
	T.~S.~Riekki, M.~S.~Manninen, J.~T.~Tuoriniemi, Phys.
	Rev. B \textbf{94} 224514 (2016).
	\newblock \doi{10.1103/physrevb.94.224514}
	
	\bibitem{Salmela_Tuoriniemi_Rysti}
	A.~Salmela, J.~Tuoriniemi, J.~Rysti, J. Low Temp. Phys.
	\textbf{162}, 678 (2010).
	\newblock \doi{10.1007/s10909-010-0246-8}
	
	\bibitem{Peshkov1975}
	V.~P.~Peshkov, JETP Lett. \textbf{21}, 162-164 (1975).
	\newblock \doi{10.1103/physrevb.39.8934}
	
	\bibitem{Murakawa2012}
	S.~Murakawa, M.~Wasai, K.~Akiyama, Y.~Wada, Y.~Tamura, R.~Nomura, Y.~Okuda, Phys. Rev. Lett. \textbf{108} 025302 (2012).
	\newblock \doi{10.1103/PhysRevLett.108.025302}

	\bibitem{Agnolet1989}
	G.~Agnolet, D.~F.~McQueeney, J.~D.~Reppy, Phys. Rev. B \textbf{39}, 8934 (1989).
	\newblock \doi{10.1103/physrevb.39.8934}
	
	\bibitem{Kim1993}
	D.~Kim, M.~Nakagawa, O.~Ishikawa, T.~Hata, T.~Kodama, Phys. Rev. Lett. \textbf{71}, 1581-1584 (1993).
	\newblock \doi{10.1103/PhysRevLett.71.1581}
		
	\bibitem{Boldarev2011}
	S.~T.~Boldarev, R.~B.~Gusev, S.~I.~Danilin, A.~Y.~Parshin, Instrum. Exp. Tech. \textbf{54}, 740
	(2011).
	\newblock \doi{10.1134/s0020441211050101}
	
	\bibitem{Sebedash1997}
	A.~P.~Sebedash, JETP Lett.
	\textbf{65}, 276 (1997).
	\newblock \doi{10.1134/1.567360}
	
	\bibitem{Adiabatic_Melting}
	A.~P.~Sebedash, J.~T.~Tuoriniemi, S.~T.~Boldarev, E.~M.~M.~Pentti, A.~J.~Salmela, J. Low Temp. Phys. \textbf{148}, 725 (2007).
	\newblock \doi{10.1007/s10909-007-9443-5}
	
	\bibitem{Tuorinemi_Martikainen_Pentti}
	J.~Tuoriniemi, J.~Martikainen, E.~Pentti, A.~Sebedash, S.~Boldarev, G.~Pickett, J. Low Temp. Phys. \textbf{129}, 531 (2002).
	\newblock \doi{10.1023/a:1021468614550}
	
	\bibitem{Pentti2006}
	E.~Pentti, J.~Tuoriniemi, A.~Salmela, A.~Sebedash, J. Low Temp. Phys. \textbf{146}, 71-83 (2006).
	\newblock \doi{10.1007/s10909-006-9267-8}
	
	\bibitem{Salmela2011}
	A.~Salmela, A.~Sebedash, J.~Rysti, E.~Pentti, J.~Tuoriniemi, Phys. Rev. B \textbf{83} 134510 (2011).
	\newblock \doi{10.1103/physrevb.83.134510}
	
	\bibitem{Riekki_Thermodynamics_adiabatic}
	T.~S.~Riekki, A.~P.~Sebedash, J.~T.~Tuoriniemi, arXiv:1810.10432 (2018)
	
	\bibitem{Sebedash_QFS}
	A.~Sebedash, S.~Boldarev, T.~Riekki, J.~Tuoriniemi, J. Low Temp. Phys.
	\textbf{187}, 588 (2017).
	\newblock \doi{10.1007/s10909-017-1755-5}
	
	
	\bibitem{Todoshchenko2014}
	I.~Todoshchenko, J.~P.~Kaikkonen, R.~Blaauwgeers, P.~J.~Hakonen, A.~Savin, Rev. Sci. Instrum.
	\textbf{85}, 085106 (2014).
	\newblock \doi{10.1063/1.4891619}
	
	\bibitem{Rollin1}
	B.~V.~Rollin, F.~Simon, Physica
	\textbf{6} 219--230 (1939).
	\newblock \doi{10.1016/S0031-8914(39)80013-1}
	
	\bibitem{Rollin2}
	H.~A.~Fairbank, C.~T.~Lane, Phys. Rev.
	\textbf{76}, 1209--1211 (1949).
	\newblock \doi{10.1103/PhysRev.76.1209}
	
	\bibitem{NematicAerogels}
	V.~E.~Asadchikov, R.~S.~Askhadullin, V.~V.~Volkov, V.~V.~Dmitriev, N.~K.~Kitaeva,
	P~.N.~Martynov, A.~A.~Osipov, A.~A.~Senin, A.~A.~Soldatov, D.~I.~Chekrygina, A.~N.~Yudin, JETP
	Lett. \textbf{101}, 556 (2015).
	\newblock \doi{10.1134/S0021364015080020}.
	
	\bibitem{PhysRevLett.120.075301}
	V.~V.~Dmitriev, A.~A.~Soldatov, A.~N.~Yudin, Phys. Rev.
	Lett. \textbf{120}, 075301 (2018).
	\newblock \doi{10.1103/PhysRevLett.120.075301}.

	
	\bibitem{PhysRevLett.117.255301}
	S.~Autti, V.~V.~Dmitriev, J.~T.~M\"akinen, A.~A.~Soldatov, G.~E.~Volovik, A.~N.~Yudin, V.~V.~Zavjalov, V.~B.~Eltsov, Phys. Rev. Lett. \textbf{117},
	255301 (2016).
	\newblock \doi{10.1103/PhysRevLett.117.255301}.
	
	
	\bibitem{Rysti2014}
	J.~T.~Tuoriniemi, M.~S.~Manninen, J.~Rysti, J. Phys. Conf. Ser. \textbf{568},
	012023 (2014).
	\newblock \doi{10.1088/1742-6596/568/1/012023}.
	
	
\end{thebibliography}

\end{document}